\documentclass[aps,pra,twocolumn,showpacs,superscriptaddress,preprintnumbers]{revtex4-2}  

\usepackage{physics}
\usepackage{graphicx} 
\usepackage{dcolumn}   
\usepackage{bm,amssymb,amsmath,dsfont,mathtools}   
\usepackage[colorlinks=true,breaklinks=true,allcolors=blue]{hyperref}
\usepackage{url}
\usepackage{txfonts}
\usepackage{float}
\usepackage{color}

\makeatletter
\let\cat@comma@active\@empty
\makeatother

\begin{document}
	
\preprint{CERN-TH-2022-030}
	
\title{Stability analysis of non-thermal fixed points in longitudinally expanding kinetic theory}
\author{Aleksandr N. Mikheev}
\affiliation{Kirchhoff-Institut f\"ur Physik,
		Ruprecht-Karls-Universit\"at Heidelberg,
		Im~Neuenheimer~Feld~227,
		69120~Heidelberg, Germany}
\affiliation{Institut f\"{u}r Theoretische Physik,
		Ruprecht-Karls-Universit\"at Heidelberg, 
		Philosophenweg 16, 
		69120 Heidelberg, Germany}
\author{Aleksas Mazeliauskas}
\affiliation{Theoretical Physics Department,
		CERN, 
		1211 Geneva 23, Switzerland}
\author{J\"{u}rgen Berges}
\affiliation{Institut f\"{u}r Theoretische Physik,
		Ruprecht-Karls-Universit\"at Heidelberg, 
		Philosophenweg 16, 
		69120 Heidelberg, Germany}
	
\date{\today}

\begin{abstract}
We use the Hamiltonian formulation of kinetic theory to perform a stability analysis of non-thermal fixed points in a non-Abelian plasma.  We construct a perturbative expansion of the Fokker-Planck collision kernel in an adiabatic approximation and show that the (next-to-)leading order solutions reproduce the known non-thermal fixed point scaling exponents. Working at next-to-leading order, we derive the stability equations for scaling exponents and find the relaxation rate to the non-thermal fixed point. This approach provides the basis for an understanding of the prescaling phenomena observed in QCD kinetic theory and non-relativistic Bose gas systems.
\end{abstract}

\maketitle
	
%======================================================================================
%======================================================================================
\section{Introduction}
\label{sec:Introduction}
Dynamics of isolated quantum many-body systems quenched far from equilibrium has been an object of intensive study during recent years. Examples range from the dynamics of quark-gluon matter created in heavy-ion collisions \cite{Berges:2013eia,Baier:2000sb} to quenches in ultracold atomic systems \cite{Polkovnikov:2010yn,Bloch:2008}. Starting from a far-from-equilibrium initial condition, these systems may exhibit a transient regime of self-similar evolution associated to a non-thermal fixed point \citep{Berges:2008wm,Schmied:2018mte}. As a result of such self-similarity, the nonequilibrium dynamics is fully encoded in a set of universal scaling exponents and functions. Over the last decade, the existence of non-thermal fixed points has been confirmed both experimentally \cite{Prufer:2018hto,Erne:2018gmz,Glidden:2020qmu} and in numerical studies \cite{PineiroOrioli:2015cpb,Karl:2016wko,Schmied:2018upn,Shen:2019jhl}. On the theoretical side, progress has been made in predicting and explaining the observed scaling exponents using various techniques such as $1/N$-resummed kinetic theory \cite{Berges:2010ez,Chantesana:2018qsb}, low-energy effective description \cite{Mikheev:2018adp}, and the functional renormalization group \cite{Mathey:2014xxa}. 

However, much less is known about how a general system evolves  to such a self-similar regime. In \cite{Schmied:2018upn} and \cite{Mazeliauskas:2018yef}, it was proposed that already before achieving fully developed scaling the system may exhibit a dramatic reduction in complexity such that its dynamics can be described by a few slowly evolving quantities. In particular, numerically solving the leading-order QCD kinetic theory \cite{Arnold:2002zm} it was observed that much before the scaling with universal exponents is established, the evolution is already governed by the fixed-point scaling function with time-dependent scaling exponents \cite{Mazeliauskas:2018yef}. In this work, we are going to consider a toy model of an expanding Yang-Mills plasma and derive approximate equations that govern the dynamics of its scaling exponents. In particular, we derive the stability equations for scaling exponents, which can be interpreted as relaxation equations to non-thermal fixed point and demonstrate for the first time that the non-thermal fixed point is stable under small perturbations.

\section{Preliminary theory}
\subsection{Prescaling}
We begin our discussion with a quick overview of the concept of (pre)scaling, in particular in the context of heavy-ion collisions. At sufficiently high energies, where the gauge coupling is small due to asymptotic freedom \cite{Gross:1973id,Politzer:1973fx}, the time evolution of gluons (g) and quarks (q) is described by distribution functions $f_{g,q}\left(\tau,p_T,p_z\right)$. Since the system is longitudinally expanding, the distributions depend on transverse ($p_T$) and longitudinal momenta ($p_z$), and on proper time ($\tau$) \cite{Bjorken:1982qr,Baym:1984np}. In the scaling regime, the original gluon distribution obeys 
\begin{equation}
\label{eq:ScalingDefinition}
f\left(\tau,p_T,p_z\right) \stackrel{\text{sclaing}}{=} \tau^{\,\alpha} f_S\left(\tau^{\,\beta}p_T,\tau^{\gamma}p_z\right),
\end{equation}
with dimensionless $\tau \to \tau/\tau_{\mathrm{ref}}$ and $p_{T,z} \to p_{T,z}/Q_s$ in terms of some (arbitrary) time $\tau_{\mathrm{ref}}$ and characteristic momentum scale $Q_s$. The exponents $\alpha$, $\beta$, and $\gamma$ are universal, and the non-thermal fixed-point distribution $f_S$ is universal up to normalizations \cite{Berges:2013fga}, which has been established numerically using classical-statistical lattice simulations \cite{Berges:2013eia}. The exponents are expected to be $\alpha_{\mathrm{BMSS}} = -2/3$, $\beta_{\mathrm{BMSS}} = 0$, and $\gamma_{\mathrm{BMSS}} = 1/3$ according to the first stage of the ``bottom-up'' thermalization scenario \cite{Baier:2000sb} based on number-conserving and small-angle scatterings, or $\alpha_{\mathrm{BD}} = -3/4$, $\beta_{\mathrm{BD}} = 0$, $\gamma_{\mathrm{BD}} = 1/4$ in a variant of bottom-up including the effects of plasma instabilities \cite{Bodeker:2005nv}.

Similarly, during prescaling the gluon distribution satisfies
\begin{equation}
\label{eq:PrescalingDefinition}
f\left(\tau,p_T,p_z\right) \stackrel{\text{presclaing}}{=} \tau^{\,\alpha(\tau)} f_S\left(\tau^{\,\beta(\tau)}p_T,\tau^{\gamma(\tau)}p_z\right),
\end{equation}
with non-universal time-dependent exponents $\alpha(\tau)$, $\beta(\tau)$, and $\gamma(\tau)$. One can therefore regard prescaling as a partial fixed point at which the scaling function $f_S$ has already reached its fixed-point form, whereas the scaling exponents $\alpha$, $\beta$, and $\gamma$ still deviate from their asymptotic values.

\subsection{Hamiltonian formulation of kinetic theory}
In order to derive equations governing the prescaling dynamics, we are going to employ the Hamiltonian formulation of kinetic theory \cite{Blaizot:2017ucy,Blaizot:2019dut,Brewer:2019oha}, the key points of which we will briefly summarize in this section. We start off with the general Boltzmann equation of a boost-invariant (in $z$-direction) and transversally homogeneous system \cite{Florkowski:2010}:
\begin{equation}
\left[\partial_{\tau} - \frac{p_z}{\tau}\partial_{p_z} \right] f\left(\tau,p_z,p_T\right) = -\mathcal{C}[f]\left(\tau,p_z,p_T\right).
\end{equation}
Here, $f$ is a distribution density, $\mathcal{C}[f]$ is a collision integral, $\tau$ is the longitudinal proper time, and $p_z$ and $p_T$ are longitudinal and traversal momenta, respectively. For the following discussion of prescaling, it will prove convenient to recast our problem into an (infinite) set of ordinary differential equations for moments of the occupation number $f$,
\begin{equation}
n_{n,m}(\tau) \equiv \int \frac{\dd[d]{\mathbf{p}}}{\left(2\pi\right)^d} p_z^{2n} p_T^m f\left(\tau,p_z,p_T\right).
\end{equation}
Although for a general collision integral the expression
\begin{equation}
\int \frac{\dd[d]{\mathbf{p}}}{\left(2\pi\right)^d} p_z^{2n} p_T^m \mathcal{C}[f]\left(\tau,p_z,p_T\right)
\end{equation}
does not have a simple form in terms of the moments $n_{n,m}$, in this work, we are going to consider the kernel that is linear in~$f$,
\begin{equation}
\label{eq:FokkerPlanckIntegral}
\mathcal{C}[f] = -\hat{q} \nabla^2_{\mathbf{p}} f,
\end{equation}
and thus allows to reformulate the problem in the form
\begin{equation}
\partial_{\log\tau} n_{n,m} = -\mathcal{H}_{n,m;n',m'} n_{n',m'},
\end{equation}
with
\begin{align}
\mathcal{H}_{n,m;n'm'}
&=
(2n + 1)\delta_{n,n'}\delta_{m,m'} - \tau\hat{q}\big[2n (2n - 1) \delta_{n-1,n'}\delta_{m,m'}\nonumber\\
&+
m^2 \delta_{n,n'}\delta_{m-2,m'}\big]\,.
\end{align}
Here, summation over repeated indices is implied and the  momentum  diffusion  parameter $\hat{q}$ is parametrically given by \cite{Kurkela:2011ti,Blaizot:2012qd}
\begin{equation}
\hat{q}(\tau) \sim \alpha_s^2 N_c^2 \int \frac{\dd[d]{\mathbf{p}}}{\left(2\pi\right)^d}  f^2\left(\tau,p_z,p_T\right)
\end{equation}
for $\mathrm{SU}(N_c)$ gauge theories in the limit of high occupancies. 
The Fokker-Planck-type collision integral \eqref{eq:FokkerPlanckIntegral} often serves as a toy model in the context of the bottom-up thermalization scenario. In the highly anisotropic limit one may, furthermore, neglect the transversal part, i.e., take $\mathcal{C}[f] = -\hat{q}\partial_{p_z}^2$, so that the ``Hamiltonian'' reduces to 
\begin{equation}
\label{eq:FokkerPlanckHamiltonian}
\mathcal{H}_{n,n;n',m'} 
= 
\left[(2n + 1)\delta_{n,n'} - q 2n (2n - 1) \delta_{n-1,n'}\right]\delta_{m,m'}
\end{equation}
and acquires a block diagonal structure. Here, for further convenience we have also introduced $q \equiv \tau\hat{q}$. 

Adopting Dirac notation we may write
\begin{equation}
\label{eq:SchrodingerMain}
\partial_y \ket{\psi} = -\hat{H}\ket{\psi}, \quad y \equiv \log\tau/\tau_0,
\end{equation}
with
\begin{equation}
n_{n,m} \equiv \bra{n,m}\ket{\psi}, \quad \mathcal{H}_{n,m;n'm'} \equiv \mel**{n,m}{\hat{H}}{n',m'},
\end{equation}
where the inner product is given by $l^2\left(\mathbb{Z}_{\geq 0} \times \mathbb{Z}_{\geq 0}\right)$ and $\lbrace\ket{n,m} \equiv \ket{n} \otimes \ket{m} \rbrace$ span the respective natural basis,
\begin{equation}
\label{eq:nm_eig}
\ket{0,0}
=
\begin{pmatrix}
1 \\
0 \\
0 \\
\vdots
\end{pmatrix}
\otimes
\begin{pmatrix}
1 \\
0 \\
0 \\
\vdots
\end{pmatrix}, \quad
\ket{1,0}
=
\begin{pmatrix}
0 \\
1 \\
0 \\
\vdots
\end{pmatrix}
\otimes
\begin{pmatrix}
1 \\
0 \\
0 \\
\vdots
\end{pmatrix},\quad \ldots
\end{equation}
Equation \eqref{eq:SchrodingerMain} with the Hamiltonian \eqref{eq:FokkerPlanckHamiltonian} will be the subject of discussion in the remaining text.

\subsection{Adiabatic approximation}
One notes that $\hat{H}$ depends on $y$ only through the parameter $q(y)$, which immediately suggests applying the well-known adiabatic approximation from quantum mechanics. In contrast to quantum mechanics (of closed systems), however, the operator $\hat{H}$ is not necessarily (anti-)~Hermitian and hence the method requires some modifications. A particularly convenient generalization to the case of non-Hermitian yet diagonalizable Hamiltonians, which we summarize in App. \ref{app:nH}, was developed in \cite{Sun:1993vw}. The key idea is to, instead, consider 
\begin{equation}
\ket{\chi(y)} \equiv U\left(q(y)\right)^{-1}\ket{\psi(y)},
\end{equation} 
with $U$ being a transformation that diagonalizes $\hat{H}$ at a given instance $y$, 
\begin{equation}
	U(q)^{-1} \hat{H}(q) U(q) = \mathrm{diag}\left(\lambda_1(q),\lambda_2(q),\ldots\right) \equiv \hat{H}_{\mathrm{d}}(q)\,.
\end{equation}
The equation for $\ket{\chi}$ is given by
\begin{equation}
\label{eq:SchrodingerChi}
\partial_y\ket{\chi} = -\hat{H}_e \ket{\chi},
\end{equation}
where
\begin{equation}
\hat{H}_e \equiv \hat{H}_d + U^{-1}\partial_y U
\end{equation}
Splitting the last term into its diagonal and off-diagonal parts,
\begin{subequations}
\begin{align}
	\hat{H}_0(q) &= \hat{H}_{\mathrm{d}}(q) + \text{diagonal part of}\left[U(q)^{-1} \partial_y U(q)\right],\\
	\hat{V}(q) &= \text{off-diagonal part of} \left[U(q)^{-1} \partial_y U(q)\right],
\end{align}	
\end{subequations}
one immediately notices that, as opposed to the diagonal piece $\hat{H}_0$, the off-diagonal term $\hat{V}$ is non-zero if and only if $\partial_y q \neq 0$. This suggests that one may treat $\hat{V}$ as a perturbation as long as $q$ depends on $y$ slowly enough and thereby construct solutions to \eqref{eq:SchrodingerChi} in a perturbative manner: 
\begin{equation}
\ket{\chi(y)} = \sum_{l=0}^{\infty} \ket{\chi^{(l)}(y)}.
\end{equation}
Here (see App.~\ref{app:nH}),
\begin{equation}
\label{eq:PsilExpansion}
	\ket{\chi^{(l)}(y)}
	=
	\sum_n C_n^{(l)}(y) \exp\left[-\int_0^y\dd{z} \epsilon_n(q(z))\right] \ket{n},
\end{equation}
with $\ket{n}$ and $\epsilon_n = \lambda_n + \partial_y \gamma_n$ being eigenvectors and eigenvalues of $\hat{H}_0$, respectively, and $\gamma_n$ being the non-Hermitian generalization of the Berry phase,
\begin{equation}
\gamma_n(y) = \int_0^y\dd{z} \bra{n} U\left(q(z)\right)^{-1} \partial_z U\left(q(z)\right)\ket{n}.
\end{equation}
The coefficients $C_n^{(l)}$ may be computed iteratively:
\begin{subequations}
\label{eq:C_iter}
\begin{align}
C_n^{(l)}(y) 
&=
-\sum_m \int_0^y \dd{z} V_{nm}(z) \exp\left[-\int_0^z\dd{s}\omega_{nm}(s) \right] C_m^{(l-1)}\left(z\right),\\
C_n^{(0)}(y)
&=
C_n^{(0)}
=
\bra{n}\ket{\chi(0)} = \mel**{n}{U\left(q(0)\right)^{-1}}{\psi(0)},
\end{align}
\end{subequations}
where
\begin{equation}
	\omega_{nm}(y) \equiv \epsilon_m(q(y)) - \epsilon_n(q(y))
\end{equation}
and
\begin{equation}
V_{nm}(y) = \bra{n} \hat{V}(q(y)) \ket{m}.
\end{equation}
For the Fokker-Planck collision kernel \eqref{eq:FokkerPlanckHamiltonian}, one has to double the number of indices: $\ket{n} \to \ket{n,m}$, cf. \eqref{eq:nm_eig}. Straightforward computations then yield (see App.~\ref{app:MatElements})
\begin{equation}
\label{eq:U_nmkl}
U_{nm,kl}(q) 
=
\begin{dcases}
\begin{pmatrix}
n \\
k
\end{pmatrix}
\frac{(2n - 1)!!}{(2k - 1)!!} q^{n-k} \delta_{ml},\quad &n \geq k, \\
0, \quad &\text{otherwise},
\end{dcases}
\end{equation}
with
\begin{equation}
U(q)^{-1} = U(-q)\,,
\end{equation}
and
\begin{equation}
\label{eq:e_and_V}
\epsilon_{nm}(q)
=
2n + 1, \quad V_{nm,kl}(q) = \partial_y q  n\left(2n - 1\right) \delta_{n,k+1} \delta_{ml}.
\end{equation}
Knowing $U^{-1}$ one may also express zeroth-order coefficients $C_{nm}^{(0)}$ in terms of initial moments of the distribution:
\begin{align}
\label{eq:C_0}
C_{nm}^{(0)}
&=
\mel**{n,m}{U\left(q(0)\right)^{-1}}{\psi(0)}\nonumber\\
&=
\sum_{k,l} \mel**{n,m}{U\left(q(0)\right)^{-1}}{k,l} \braket{k,l\,}{\psi(0)\vphantom{U\left(q(0)\right)^{-1}}}\nonumber\\
&=
\sum_{k=0}^n \binom{n}{k} \frac{(2n - 1)!!}{(2k - 1)!!} \left(-q(\tau_0)\right)^{n-k} n_{k,m}(\tau_0). 
\end{align}

\section{Prescaling}
Now we are ready to study the time-dependent scaling exponents (prescaling)
in the adiabatic approximation of the Hamiltonian formalism. To make analytical progress, we will define a small expansion parameter and will study the scaling exponents' behavior at leading and next-to-leading orders.

First, consider the $l$-th order contribution to the $(n,m)$-th moment of the distribution function:
\begin{align}
	\bra{n,m}\ket{\psi^{(l)}(y)} 
	&=
	\sum_{kp} C_{kp}^{(l)}(y)\,\mathrm{e}^{-(2k + 1)y} \bra{n,m}U(q)\ket{k,p}\nonumber\\
	&=
	\sum_{k=0}^n  \binom{n}{k} \frac{(2n - 1)!!}{(2k - 1)!!} C_{km}^{(l)}\left(y\right) \mathrm{e}^{(n-3k-1)y} \hat{q}^{n-k}.
\end{align}
Here, we have used \eqref{eq:U_nmkl} to get the second line. For the perturbation \eqref{eq:e_and_V} the iterative relation \eqref{eq:C_iter} takes a very simple form:
\begin{align}
\label{eq:iter}
	C_{km}^{(l)}(y) 
	&=
	-\sum_{pr}\int_0^y\dd{z} V_{km,pr}(z) \exp\left[-\int_0^z\dd{s} \omega_{km,pr}(s)\right] C_{pr}^{(l-1)}(z)\nonumber\\
	&=
	-k(2k - 1)\int_0^y\dd{z} \partial_z q\left(z\right)\mathrm{e}^{2z} C_{k-1 m }^{(l-1)}(z)\,. 
\end{align}
Upon repeating the procedure \eqref{eq:iter} $l$ times one readily obtains
\begin{align}
	&C_{km}^{(l)}(y)
	=
	C_{k-l m}^{(0)} (-1)^l \frac{k!}{(k-l)!} \frac{(2k - 1)!!}{(2k-2l - 1)!!}\nonumber\\
	&\times \int_0^y \dd{z_1} \int_0^{z_1} \dd{z_2} \ldots \int_0^{z_{l-1}} \dd{z_l} \mathcal{V}(z_1) \mathcal{V}(z_2) \ldots \mathcal{V}(z_l)\,,
\end{align}
with $C^{(l > k)}_{km} \equiv 0$ and $\mathcal{V}(z) \equiv \exp(2z)\, \partial_z q(z)$. We now recall that
\begin{align}
	&\phantom{= =}\int_0^y \dd{z_1} \int_0^{z_1} \dd{z_2} \ldots \int_0^{z_{l-1}} \dd{z_l} \mathcal{V}(z_1) \mathcal{V}(z_2) \ldots \mathcal{V}(z_l)\nonumber\\
	&=
	\int_0^y \dd{z_1} \int_0^{z_1} \dd{z_2} \ldots \int_0^{z_{l-1}} \dd{z_l} \mathcal{T} \left\lbrace\mathcal{V}(z_1) \mathcal{V}(z_2) \ldots \mathcal{V}(z_l)\right\rbrace\nonumber\\
	&=
	\frac{1}{l!} \int_0^y \dd{z_1} \int_0^y \dd{z_2} \ldots \int_0^y \dd{z_l} \mathcal{T} \left\lbrace\mathcal{V}(z_1) \mathcal{V}(z_2) \ldots \mathcal{V}(z_l)\right\rbrace,
\end{align}
where $\mathcal{T}$ is a time-ordering operator. Since $\mathcal{V}(z_i)$ are ordinary numbers, the time-ordering operator drops out and we are left with
\begin{equation}
C_{km}^{(l)}(\tau)
=
(-1)^l C_{k-l m}^{(0)}  
\begin{pmatrix}
k\\
l
\end{pmatrix}	
\frac{(2k - 1)!!}{(2k-2l - 1)!!} \left[\frac{v(\tau)}{\Delta(\tau)}\right]^l,
\end{equation}
where we went back to $\tau = \tau_0 \exp(y)$ (setting also $\tau_0=1$ for brevity). Here, we have also introduced the functions
\begin{align}
v(\tau)
&=
\frac{1}{\tau^3\hat{q}(\tau)} \int_1^{\tau} \dd{\tau'} \left(\tau'\right)^2 \pdv{q(\tau')}{\tau'}\nonumber\\
&=
\frac{1}{\tau^3\hat{q}(\tau)} \int_1^{\tau} \dd{\tau'} \left(\tau'\right)^2 \hat{q}(\tau')\left[1 + \pdv{\log\hat{q}(\tau')}{\log\tau'}\right]
\end{align}
and 
\begin{equation}
\label{eq:tau^3q}
\Delta(\tau) =\frac{1}{\tau^3 \hat{q}(\tau)}
\end{equation}
that will serve as an expansion parameter. Assembling everything together we end up with
\begin{align}
\label{eq:psi_l_gen}
\bra{n,m}\ket{\psi^{(l)}(\tau)} 
&=
(2n - 1)!! n! \tau^{n-1} \left[\hat{q}(\tau)\right]^n  \left[-v(\tau)\right]^l \nonumber\\
& \hspace{-30pt} \times
\sum_{k=l}^n \frac{C_{k-l m}^{(0)}}{l!(k-l)!(n-k)!(2k - 2l - 1)!!}  \left[\Delta(\tau)\right]^{k - l}.
\end{align}

\subsection{Perturbative expansion}
\begin{figure}	
\includegraphics[width=0.45\textwidth]{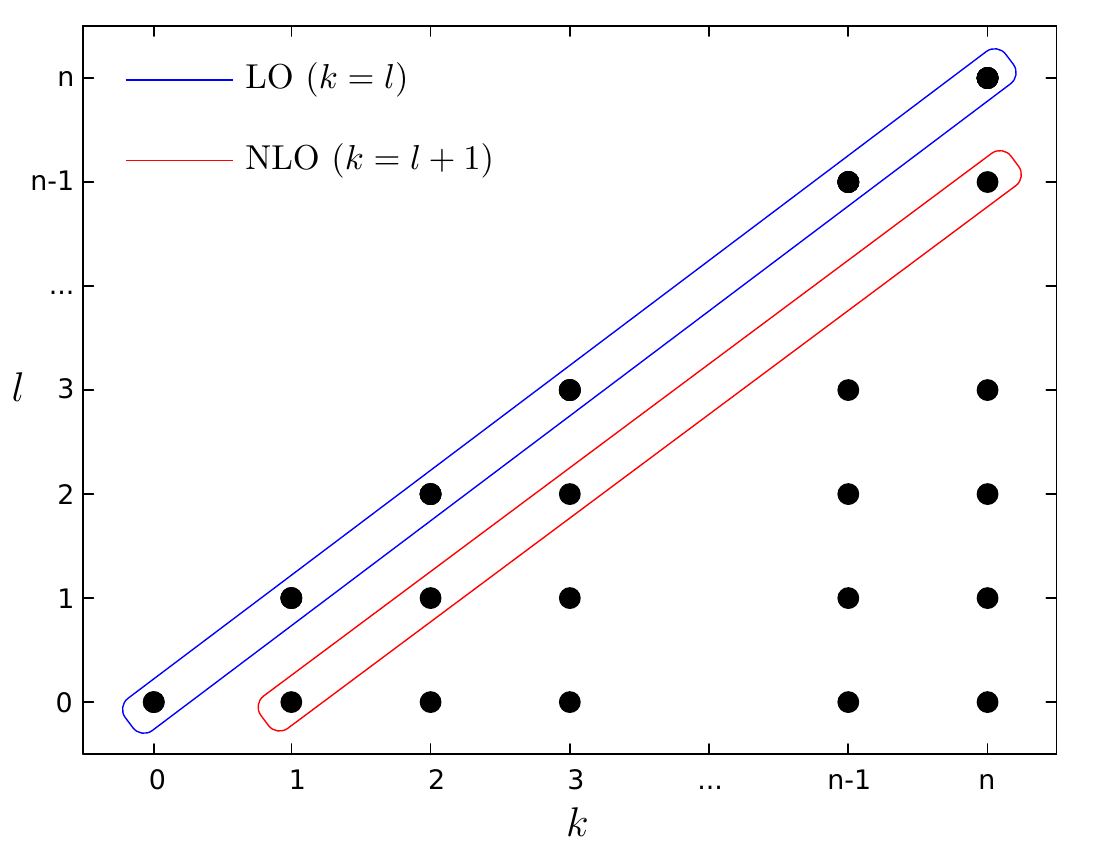}
\caption{Schematic depiction of the resummation scheme. At any order $l$ of perturbation theory the leading-order (blue) and the next-to-leading-order (red) contributions to $\bra{n,m}\ket{\psi}$ are given by the $l$-th and $(l+1)$-st term in the expansion \eqref{eq:psi_l_gen}, respectively. In both cases the resulting behavior does not depend on the order $l$, cf. \eqref{eq:LO} and \eqref{eq:NLO}, so that one has to take all the orders of perturbation theory into account.}
\label{fig:domain}
\end{figure}

To go further, we need to truncate the series \eqref{eq:psi_l_gen}. To do so, let us first estimate the large-time behavior of the quantities entering the expansion \eqref{eq:psi_l_gen}. Near the fixed point, $\hat{q} \sim \tau^{2\alpha_{*} - 2\beta_{*} - \gamma_{*}}$ implying the large-time behavior
\begin{equation}
\Delta(\tau \gg 1) \sim \tau^{-2\alpha_{*} + 2\beta_{*} + \gamma_{*} - 3}.
\end{equation}
Hence, if $2\alpha_{*} - 2\beta_{*} - \gamma_{*} + 3 > 0 $, then we expect $\Delta(\tau)$ to decay at large times $\tau$ and therefore may use it as a small parameter, at least when the scaling exponents are not too far off from their asymptotic values. Note that this condition holds both for the bottom-up \cite{Baier:2000sb} and for the modified \cite{Bodeker:2005nv} scaling solutions. On the contrary, for $v$ the same analysis results in
\begin{equation}
v(\tau \gg 1) \sim \mathrm{const}.
\end{equation}
One may even estimate the asymptotic value as
\begin{equation}
\label{eq:v_asymp}
v(\tau \to \infty) = \frac{1 + 2\alpha_{*} - 2\beta_{*} - \gamma{*}}{3 + 2\alpha_{*} - 2\beta_{*} - \gamma_{*}}.
\end{equation}
We thus conclude that the $k$-th term in $\bra{n,m}\ket{\psi^{(l)}(\tau)}$ at large times scales as
\begin{equation}
\bra{n,m}\ket{\psi^{(l)}(\tau \gg 1)}_{k\text{-th term}} \sim \tau^{n - 1} \hat{q}^n  \Delta^{k - l}, \quad n \geq k \geq l,
\end{equation}
with $\Delta(\tau)$ playing a role of the small parameter. The leading order (LO) contribution to $\bra{n,m}\ket{\psi^{(l)}(y)}$ is then given by the $l$-th term in \eqref{eq:psi_l_gen},
\begin{equation}
\label{eq:LO}
	\bra{n,m}\ket{\psi^{(l)}(\tau)}_{\text{LO}} \sim \tau^{n - 1} \hat{q}^n, \quad n \geq  l,
\end{equation}
the next-to-leading order (NLO) contribution is given by the $(l+1)$-st term,
\begin{equation}
\label{eq:NLO}
	\bra{n,m}\ket{\psi^{(l)}(\tau)}_{\text{NLO}} \sim \tau^{n - 1} \hat{q}^n \Delta\,, \quad n \geq  l + 1,
\end{equation}
etc. Importantly, this behavior is independent of $l$, which alludes to a possible need of resummation of all the terms of the same kind: 
\begin{align}
\bra{n,m}\ket{\psi(\tau)} 
&=
\underbrace{\sum_{l=0}^n \bra{n,m}\ket{\psi^{(l)}(\tau)}_{l\text{-th term}}}_{\mathrm{LO} \iff O(\Delta^0)}\nonumber\\ 
&+
\underbrace{\sum_{l=0}^{n-1} \bra{n,m}\ket{\psi^{(l)}(\tau)}_{(l+1)\text{-st term}}}_{\mathrm{NLO} \iff O(\Delta^1)} + \ldots,
\end{align}
see Fig.~\ref{fig:domain} for visualization and more details. 

We are now in a position to derive equations that govern the prescaling dynamics at next-to-leading order. According to the above discussion, at this order the $(n,m)$-th moment of the distribution takes the form
\begin{align}
\bra{n,m}\ket{\psi(\tau)}_{\mathrm{NLO}}
&=
(2n - 1)!! \tau^{n-1} \hat{q}(\tau)^n \,\Bigg\lbrace C_{0 m}^{(0)} \sum_{l=0}^n \binom{n}{l} \left[-v(\tau)\right]^l\nonumber\\
&+
C_{1 m}^{(0)} n \Delta(\tau) \sum_{l=0}^{n-1} \binom{n-1}{l} \left[-v(\tau)\right]^l \Bigg\rbrace\,.
\end{align}
Recognizing the binomial expansion we readily obtain
\begin{equation}
\label{eq:nlo}
\bra{n,m}\ket{\psi(\tau)}_{\mathrm{NLO}}
=
a_{nm} \tau^{n-1} \hat{q}(\tau)^n \left[1 - v(\tau)\right]^n \left\lbrace 1 + b_m \frac{n \Delta(\tau)}{1 - v(\tau)}  \right\rbrace,
\end{equation}
with $a_{nm} = C_{0m}^{(0)} (2n - 1)!!$ and $b_m = C_{1m}^{(0)}/C_{0m}^{(0)}$.

\subsection{Fixed-point equations}
Up until this point, we have not assumed any particular ansatz for the time evolution of moments $n_{n,m}$ of the distribution function. If the prescaling assumption \eqref{eq:PrescalingDefinition} holds, however, then we can recast the equations for the moments in terms of time-dependent scaling exponents. 
Following the original work \cite{Mazeliauskas:2018yef}, in order to reflect instantaneous scaling properties we redefine exponents in \eqref{eq:PrescalingDefinition} as
\begin{equation}
	\tau^{\alpha(\tau)} \to \exp\left[\int_1^{\tau} \frac{\dd{\tau'}}{\tau'} \alpha\left(\tau'\right)\right],
\end{equation}
which for constant $\alpha$ reduces to the power law $\tau^{\alpha}$. The rate of change of a particular moment $n_{n,m}$ as well as of the momentum diffusion parameter $\hat{q}$ is given by a linear combination of scaling exponents:
\begin{subequations}
	\label{eq:n_and_q_presc}	
	\begin{align}
		\label{eq:dn_presc}	
		\pdv{\log n_{n,m}(\tau)}{\log\tau} 
		&=
		\alpha(\tau) - \left(m + 2\right) \beta(\tau) - \left(2n + 1\right) \gamma(\tau)\,,\\
		\label{eq:dlogq}
		\pdv{\log \hat{q}(\tau)}{\log\tau} 
		&=
		2 \alpha(\tau) - 2\beta(\tau) - \gamma(\tau)\,,
	\end{align}
\end{subequations}
in $d=3$ spatial dimensions. This also implies
\begin{subequations}
\begin{align}
\label{eq:dv_presc}
\pdv{v(\tau)}{\log\tau}
&=
\left[1 + 2\alpha(\tau) - 2 \beta(\tau) - \gamma(\tau)\right]\left[1 - v(\tau)\right]-2 v(\tau)\,,\\
\label{eq:dlogD_presc}
\pdv{\log \Delta(\tau)}{\log\tau}
&=
-3 - 2\alpha(\tau) + 2\beta(\tau) + \gamma(\tau)\,.
\end{align}
\end{subequations}
Taking then the log of both sides of \eqref{eq:nlo} and then the derivative with respect to $\log \tau$ we end up with
\begin{align}
\alpha &- \left(m + 2\right)\beta - \left(2n+1\right)\gamma 
=
n - 1 + n\left(2\alpha - 2\beta - \gamma\right) + n\frac{2 v}{1 - v}\nonumber\\ 
&-
n\left(1 + 2\alpha - 2\beta - \gamma\right) 
+
\pdv{}{\log \tau} \log \left(1 + b_m \frac{n\Delta}{1 - v}  \right).
\end{align}	
Since the NLO approximation is $O(\Delta)$, we have to expand the log term on the right-hand side to first order in $\Delta$ to be consistent. After some simple algebra, one then eventually arrives at
\begin{align}
\label{eq:pre_floweq}
\alpha - 2\beta - \gamma + 1 - m \beta - 2 n \left[\gamma + \frac{v}{1-v} - \frac{b_m \Delta}{(1 - v)^2} \right] = 0\,.
\end{align}	
First, we observe that in order for this equation to hold for any $n$ and $m$ (as it should during prescaling) one has to impose
\begin{equation}
\label{eq:NumberConservation}
\alpha(\tau) - 2\beta(\tau) - \gamma(\tau) + 1 = 0\,.
\end{equation}
One immediately recognizes in \eqref{eq:NumberConservation} the scaling relation that follows from conservation of the total particle number \cite{Berges:2013fga}. This reflects the particle-number-conserving nature of the elastic collision kernel. It is then suggestive to also demand that the term containing $m$ and the term containing $n$ should individually vanish identically, too. This would result in another constraint
\begin{equation}
\label{eq:EnergyConservation}
\beta(\tau) = 0\,,
\end{equation}
which together with \eqref{eq:NumberConservation} indicates energy conservation \cite{Berges:2013fga}. The remaining equation then reads
\begin{equation}
\label{eq:final_pre_floweq}
\gamma + \frac{v}{1 - v} = \frac{b_m \Delta}{(1 - v)^2}\,.
\end{equation}
For this condition to hold $b_m$ has to be $m$-independent. Since
\begin{equation}
	b_m = \frac{C_{1m}^{(0)}}{C_{0m}^{(0)}} = 1 - q(\tau_0)\,\frac{n_{1m}(\tau_0)}{n_{0m}(\tau_0)}\,,
\end{equation}
see \eqref{eq:C_0}, the latter holds as long as $n_{1m}(\tau_0)/n_{0m}(\tau_0)$ does not depend on $m$. An important class of distributions for which this condition is always satisfied is given by separable distributions, i.e., $f(\tau_0,p_T,p_z) = f_1(\tau_0,p_T)\,f_2(\tau_0,p_z)$.

\begin{figure}	
	\includegraphics[width=0.45\textwidth]{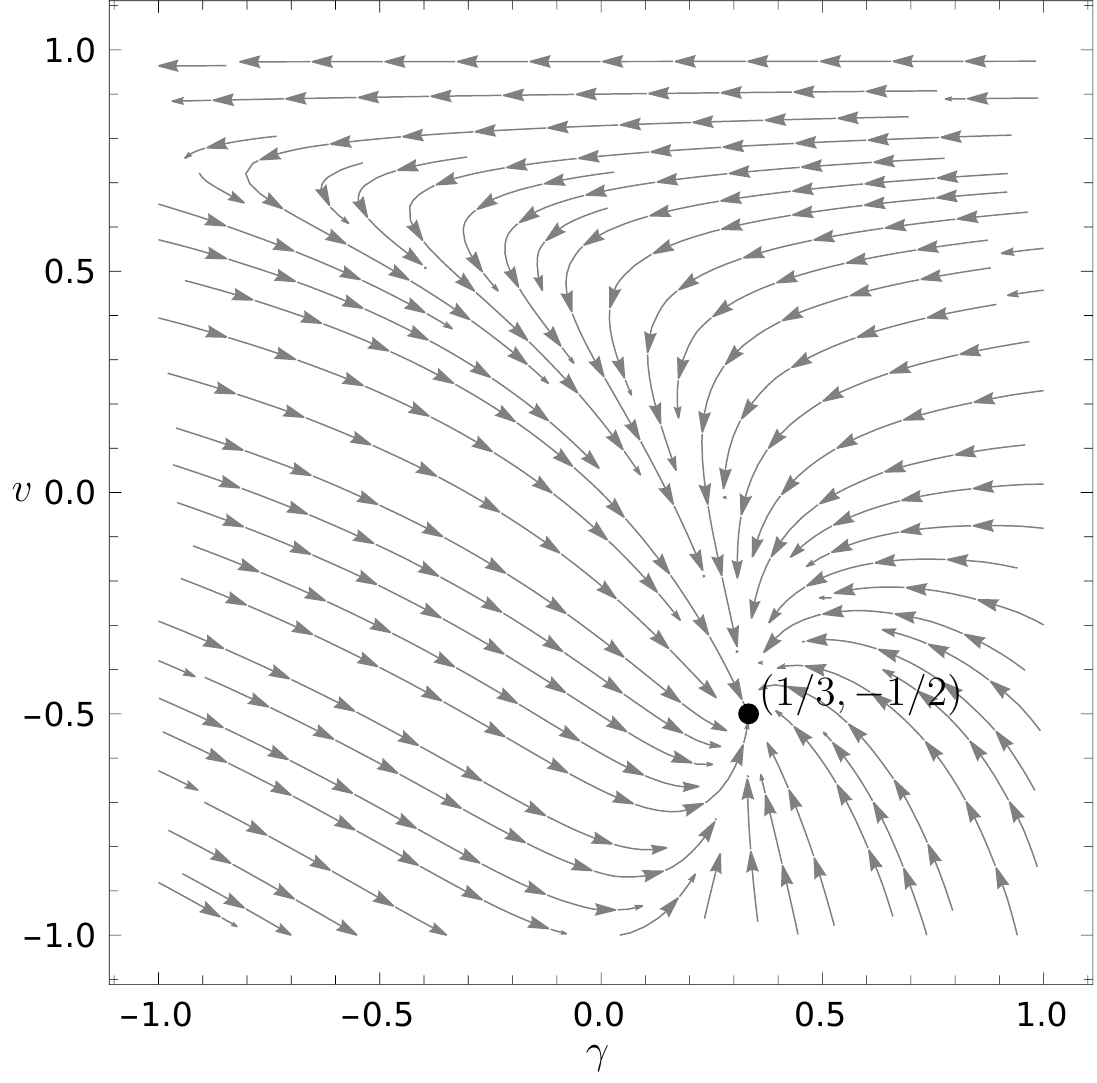}
	\caption{The flow diagram of \eqref{eq:betas}. The sole fixed point $(1/3,-1/2)$ corresponds to the ``bottom-up'' scaling \cite{Baier:2000sb}.}
	\label{fig:flow}
\end{figure}

We have already derived the equation \eqref{eq:dv_presc} governing the dynamics of $v$ during prescaling. To obtain a similar equation for the remaining scaling exponent $\gamma$, we first take one more logarithmic derivative of both sides of \eqref{eq:final_pre_floweq}:
\begin{equation}
\frac{\dot{\gamma} + \dot{v}/(1-v)^2}{\gamma + v/(1-v)} = \pdv{\log\Delta}{\log\tau} + \frac{2 \dot{v}}{1-v},
\end{equation}
where $\dot{\mathcal{O}} \equiv \partial_{\log\tau} \mathcal{O}$. Finally, using \eqref{eq:dv_presc} and \eqref{eq:dlogD_presc} and imposing the constraints \eqref{eq:NumberConservation} and \eqref{eq:EnergyConservation} one ends up with the system of differential equations:
\begin{equation}
\label{eq:FlowEquations}
\dot{\boldsymbol{\kappa}}
=
\boldsymbol{\mathcal{B}}\left(\boldsymbol{\kappa}\right),
\end{equation}
where we have introduced $\boldsymbol{\kappa} = \left(\gamma,v\right)$ and
\begin{subequations}
\label{eq:betas}
\begin{align}
\mathcal{B}_{\gamma}(\boldsymbol{\kappa})
&=
-\dfrac{v^2 + 2v - 1 - \left(1 - v\right)^2 \gamma^2 + 4\left(1 - v\right)\gamma}{(1 - v)^2},\\
\mathcal{B}_v(\boldsymbol{\kappa})
&=
\gamma - 1 - \left(\gamma + 1\right) v\,.
\end{align}
\end{subequations}
The above equations resemble flow equations describing a running of couplings in the context of renormalization group flow. The flow diagram of \eqref{eq:betas} is depicted in Fig.~\ref{fig:flow}.

Scaling is achieved when the flow reaches a fixed point
\begin{equation}
\boldsymbol{\mathcal{B}}\left(\boldsymbol{\kappa}_{*}\right) \stackrel{\text{scaling}}{=} 0. 
\end{equation}
Using \eqref{eq:betas} one recognizes the standard bottom-up scaling exponents,
\begin{equation}
\alpha_{*} = -2/3,\quad \beta_{*} = 0, \quad \gamma_{*} = 1/3, 
\end{equation}
together with 
\begin{equation}
v_{*} = -1/2,
\end{equation}
cf. \eqref{eq:v_asymp}, as a stable fixed point of the flow equations \eqref{eq:FlowEquations}. Indeed, using the standard notation $\delta \boldsymbol{\kappa} \equiv \boldsymbol{\kappa} - \boldsymbol{\kappa}_{*}$ one has
\begin{equation}
\delta\dot{\boldsymbol{\kappa}} 
\stackrel{\text{bottom-up}}{=}
\begin{pmatrix}
-2 & 0  \\
3/2 & -4/3 
\end{pmatrix}
\delta\boldsymbol{\kappa}
+
O\left(\delta\kappa_i^2\right).
\end{equation}
The corresponding characteristic polynomial reads $\left(\lambda + 2\right)\left(\lambda + 4/3\right)$ resulting in two (simple) eigenvalues
\begin{equation}
	\lambda_1 = -2, \quad \lambda_2 = -4/3,
\end{equation}
with the respective eigenvectors
\begin{equation}
\mathbf{h}_1 = \left(-4/9,1\right)^T, \quad \mathbf{h}_2 = \left(0,1\right)^T.
\end{equation}	
The general solution near the fixed point is therefore given by
\begin{equation}
\delta\boldsymbol{\kappa}
=
K_1 \tau^{\,\lambda_1} \mathbf{h}_1 + K_2 \tau^{\,\lambda_2} \mathbf{h}_2,
\end{equation}
or explicitly,
\begin{subequations}
\begin{align}
\delta \gamma(\tau) 
&=
-\frac{4}{9} K_1 \tau^{-2},\\
\delta v(\tau)
&=
K_1 \tau^{-2} + K_2 \tau^{-4/3}\,.
\end{align}
\end{subequations}

\section{Conclusions}
In this work we studied the self-similar evolution phenomena in Fokker-Planck type kinetic theory. Using the Hamiltonian formalism of kinetic theory and adiabatic approximation, we were able to derive the flow equations of the time-dependent scaling exponents. The fixed point of scaling exponents for the Fokker-Planck kinetic theory coincides with the scaling exponents characterizing the early stage of the bottom-up thermalization scenario \cite{Baier:2000sb}.

Working at next-to-leading order in the small expansion parameter, we found the relaxation rate for scaling exponents to the fixed point and demonstrated its stability. This analysis lays ground for the study of scaling phenomena in more complex systems, such as full QCD kinetic theory. 

\begin{center}$ {\ast}\,{\ast}\,{\ast} $\end{center}

We note that an analysis of time-dependent scaling exponents in Fokker-Planck kinetic theory was performed independently by Jasmine Brewer, Bruno Scheihing-Hitschfeld and Yi Yin and made public simultaneously to the present manuscript \cite{Brewer:2022vkq}.

%======================================================================================
%======================================================================================
\begin{acknowledgments}
The authors thank Jasmine Brewer, Bruno Scheihing-Hitschfeld and Yi Yin for useful discussions. In particular, the authors acknowledge the presentation by  Bruno Scheihing-Hitschfeld at Initial Stages Conference 2021, which motivated the present work. This work is funded by Deutsche Forschungsgemeinschaft (DFG, German Research Foundation) under SFB 1225 ISOQUANT (project ID 27381115) and under Germany’s Excellence Strategy EXC2181/1-390900948 -- the Heidelberg STRUCTURES Excellence Cluster. ANM acknowledges financial support by the IMPRS-QD (International Max Planck Research School for Quantum Dynamics).
\end{acknowledgments}
	
\bibliographystyle{apsrev4-2}
\bibliography{references}	
	
\newpage	
	
%======================================================================================
%======================================================================================

\begin{appendix}
\begin{center}
\textbf{APPENDIX}
\end{center}
\setcounter{equation}{0}
\setcounter{table}{0}
\makeatletter
\section{Non-Hermitian adiabatic expansion}
\label{app:nH}

In this appendix, we show how one can systematically solve equations of the kind
\begin{equation}
	\label{eq:Schrodinger}
	\partial_y \ket{\psi(y)} = - \hat{H}(q(y)) \ket{\psi(y)},
\end{equation}
where $\hat{H}$ is diagonalizable but not necessarily \mbox{(anti-)Hermitian}. Following \cite{Sun:1993vw}, we first want to find a transformation $U$ that diagonalizes $\hat{H}$ at each instance $y$,
\begin{equation}
	U(q)^{-1} \hat{H}(q) U(q) = \mathrm{diag}\left(\lambda_1(q),\lambda_2(q),\ldots\right) \equiv \hat{H}_{\mathrm{d}}(q).
\end{equation}
For example, in the standard basis $\lbrace\ket{k}\rbrace$ this transformation reads
\begin{equation}
	U(q) = \sum_k \ket{v_k(q)}\bra{k}, \quad \hat{H}(q)\ket{v_k(q)} = \lambda_k(q) \ket{v_k(q)}.
\end{equation}
Let $\ket{\psi}$ be a solution to the equation \eqref{eq:Schrodinger}. Define the ``equivalent solution'',
\begin{equation}
	\ket{\chi(y)} \equiv U(q(y))^{-1}\ket{\psi(y)},
\end{equation}
that satisfies 
\begin{equation}
	\partial_y \ket{\chi(y)} = -\hat{H}_{\mathrm{e}}(q(y))\ket{\chi(y)},
\end{equation}
with
\begin{equation}
	\hat{H}_{\mathrm{e}}(q) = \hat{H}_{\mathrm{d}}(q) + U(q)^{-1} \partial_y U(q) .
\end{equation}
We now split the second term into diagonal and off-diagonal parts and introduce
\begin{equation}
	\hat{H}_0(q) = \hat{H}_{\mathrm{d}}(q) + \text{diagonal part of}\left[U(q)^{-1} \partial_y U(q)\right]
\end{equation}
and
\begin{equation}
	\hat{V}(q) = \text{off-diagonal part of} \left[U(q)^{-1} \partial_y U(q)\right],	
\end{equation}
so that
\begin{equation}
	\hat{H}_{\mathrm{e}}(q) = \hat{H}_0(q) + \hat{V}(q).
\end{equation}
One can already guess that the diagonal part $\hat{H}_0(q)$ governs adiabatic element of the evolution, whereas the off-diagonal piece $\hat{V}(q)$ gives rise to non-adiabatic transitions between the quasi-energy levels. Furthermore, since $\hat{V}(q)$ vanishes when there is no time-dependence we anticipate that one can treat $\hat{V}(q)$ as a perturbation when $q(y)$ depends on $y$ slowly enough. We will therefore look for solutions in the form 
\begin{equation}
	\ket{\chi(y)} = \sum_{l=0}^{\infty} \ket{\chi^{(l)}(y)}.
\end{equation} 
Here, 
\begin{subequations}
	\label{eq:pert_eqs}	
	\begin{align}
		\partial_y \ket{\chi^{(0)}(y)} 
		&=
		-\hat{H}_0(q(y)) \ket{\chi^{(0)}(y)},\\
		\partial_y \ket{\chi^{(l)}(y)}
		&=
		-\hat{H}_0(q(y)) \ket{\chi^{(l)}(y)} - \hat{V}(q) \ket{\chi^{(l-1)}(y)},
	\end{align}
\end{subequations}
for $l \geq 1$. The zeroth order solution is given by
\begin{equation}
	\ket{\chi^{(0)}(y)} = \exp\left[-\int_0^y \dd{z} \hat{H}_0(q(z)) \right] \ket{\chi(0)}, 
\end{equation}
with $\ket{\chi(0)} \equiv U(q(0))^{-1} \ket{\psi(0)}$. It is convenient to work in the basis of eigenvectors $\hat{H}_0$, which we can choose to be
\begin{equation}
	\ket{0} 
	=
	\begin{pmatrix}
		1 \\
		0 \\
		0 \\
		\vdots
	\end{pmatrix},
	\quad
	\ket{1} 
	=
	\begin{pmatrix}
		0 \\
		1 \\
		0 \\
		\vdots
	\end{pmatrix},	
	\quad 
	\ldots \quad.
\end{equation}
The corresponding eigenvalues read $\epsilon_n = \lambda_n + \partial_y\gamma_n$, with
\begin{equation}
	\gamma_n(y)
	=
	\int_0^y \dd{z} \bra{n} U(q(z))^{-1} \partial_z U(q(z)) \ket{n}.
\end{equation}
Expanding the $l$-th order solution in this basis as
\begin{equation}
	\ket{\chi^{(l)}(y)}
	=
	\sum_n C_n^{(l)}(y) \exp\left[-\int_0^y\dd{z} \epsilon_n(q(z))\right] \ket{n}
\end{equation}
and substituting it into \eqref{eq:pert_eqs} we obtain, for $l\geq1$,
\begin{align}
	&\phantom{-}
	\sum_m \partial_y C^{(l)}_m(y)\, \exp\left[-\int_0^y\dd{z} \epsilon_m(q(z))\right] \ket{m}\nonumber\\
	&-
	\sum_m C^{(l)}_m(y)\, \epsilon_m(y) \exp\left[-\int_0^y\dd{z} \epsilon_m(q(z))\right] \ket{m}\nonumber\\
	=
	&-
	\sum_m C^{(l)}_m(y)\, \epsilon_m(q(y)) \exp\left[-\int_0^y\dd{z} \epsilon_m(q(z))\right] \ket{m}\nonumber\\
	&-
	\sum_m C^{(l-1)}_m(y) \exp\left[-\int_0^y\dd{z} \epsilon_m(q(z))\right] \hat{V}(q(y))\ket{m},
\end{align}
where we have used $\hat{H}_0\ket{m} = \epsilon_m\ket{m}$. First, we notice that the last term on the left-hand side cancels the first term on the right-hand side. Multiplying both sides by $\bra{n}$ and using the orthogonality condition $\bra{n}\ket{m} =\delta_{mn}$ we then readily obtain
\begin{equation}
	\partial_y C_n^{(l)}(y) 
	=
	-\sum_m V_{nm}(y) \exp\left[-\int_0^y\dd{s}\omega_{nm}(s) \right] C_m^{(l-1)}(y),
\end{equation}
with
\begin{equation}
	\omega_{nm}(y) \equiv \epsilon_m(q(y)) - \epsilon_n(q(y))
\end{equation}
and 
\begin{equation}
	V_{nm}(y) \equiv \bra{n}\hat{V}(q(y))\ket{m}.
\end{equation}
We thus conclude
\begin{equation}
	\label{eq:c_n^l}
	C_n^{(l)}(y) 
	=
	-\sum_m \int_0^y \dd{z} V_{nm}(z) \exp\left[-\int_0^z\dd{s}\omega_{nm}(s) \right] C_m^{(l-1)}(z).
\end{equation}
As a final remark, we note that tedious, yet straightforward computations show that there is also no ambiguity regarding the choice of the instantaneous eigenfunctions $\ket{v_k(q)}$. In other words, $\ket{\psi^{(l)}}$ are invariant under reparameterizations
\begin{equation}
	\ket{v_k(y)} \to \mathrm{e}^{\phi_k(y)} \ket{v_k(y)}
\end{equation}
at each order of perturbation theory.

\section{Computation of $U$, $\hat{H}_0$, and $\hat{V}$}
\label{app:MatElements}

\subsection{Solving eigenproblem}
To find a transformation $U$ that diagonalizes the matrix \eqref{eq:FokkerPlanckHamiltonian}, one shall solve the corresponding eigenproblem.
\begin{equation}
\hat{H}(q) \ket{v_{k,l}} = \lambda_{k,l} \ket{v_{k,l}}.
\end{equation}
Here, the subscript enumerates eigenvalues and eigenvectors. Since $\hat{H}$ has a block diagonal structure, it obviously suffices to study only one block as generalization to the full case is straightforward. In Dirac notation,
\begin{equation}
\label{eq:eigenproblem}
\sum_{n'} \mel**{n}{\hat{H}(q)}{n'} \bra{n'}\ket{v_k} = \lambda_k \bra{n}\ket{v_k},
\end{equation}
with
\begin{equation}
\bra{n} \hat{H}(q) \ket{n'}
=
\left(2 n + 1\right) \delta_{n,n'} - q \,2n\left(2n - 1\right) \delta_{n-1,n'}\,.
\end{equation}
Since $\hat{H} - \lambda \hat{I}$ is bidiagonal and determinant of a bidiagonal matrix is equal to product of its diagonal elements, the characteristic equation simply reads
\begin{equation}
\prod_{n\geq0}\left(2 k + 1 - \lambda\right) = 0,
\end{equation}
from which we easily deduce
\begin{equation}
\lambda_k = 2k + 1.
\end{equation}
Plugging this into \eqref{eq:eigenproblem} yields the recursion relation
\begin{equation}
\bra{n-1}\ket{v_k} = \frac{n - k}{q n (2n - 1)} \bra{n}\ket{v_k}\,, \quad n \geq 1.
\end{equation}
One can verify that $\bra{n<k}\ket{v_k} = 0$. It is then suggestive to set $\bra{k}\ket{v_k} = 1$ and compute the remaining components of each eigenvector ascending with
\begin{equation}
\bra{n}\ket{v_k} = q\,\frac{n (2n - 1)}{n - k} \bra{n-1}\ket{v_k}\,, \quad n > k.
\end{equation}
Hence,
\begin{equation}
\bra{n}\ket{v_k} = \prod_{p > k}^n \frac{p}{p - k} (2p - 1)\,q 
=
\begin{pmatrix}
n \\ 
k
\end{pmatrix} 
\frac{(2n - 1)!!}{(2k - 1)!!} \,q^{n-k}\,, \quad n > k.
\end{equation}
Here, we have used
\begin{equation}
\prod_{p>k}^n \frac{p}{p - k} = \frac{(k + 1)\cdot(k + 2)\cdot \ldots \cdot n}{1 \cdot 2 \cdot \ldots \cdot (n - k)} = \frac{n!}{k!(n - k)!} = \begin{pmatrix}
n \\ 
k
\end{pmatrix}
\end{equation}
and
\begin{equation}
\prod_{p > k}^n (2p - 1) = \frac{\prod_{p = 1}^n (2p - 1)}{\prod_{p=1}^k (2 p - 1)} = \frac{(2n - 1)!!}{(2k - 1)!!}\,,
\end{equation}
with the standard convention $(-1)!! = 1$. We therefore conclude
\begin{equation}
\label{eq:U_nm}
U_{nm}(q) 
=
\begin{dcases}
\begin{pmatrix}
n \\
m
\end{pmatrix}
\frac{(2n - 1)!!}{(2m - 1)!!} q^{n-m},\quad &n \geq m, \\
0, \quad &\text{otherwise}.
\end{dcases}
\end{equation}

\subsection{Finding $U(q)^{-1}$}
For brevity, we are going to temporarily denote the entries of $U(q)$ and $U(q)^{-1}$ by $a_{nm}$ and $b_{nm}$, respectively. Let us prove that 
\begin{equation}
U(q)^{-1} = U(-q) \iff b_{nm} = (-1)^{n-m} a_{nm}.
\end{equation}
We are going to do so by induction. First, we note that since $U(q)$ is lower triangular, $U(q)^{-1}$ is also lower triangular. The diagonal elements of a product are then just a product of diagonal elements, which implies $
1 = a_{nn} b_{nn}  = b_{nn}$. The second row yields, on top of that, one non-trivial condition:
\begin{equation}
0 = a_{10} b_{00} + a_{11} b_{10} = a_{10} + b_{10} \implies b_{10} = -a_{10}. 
\end{equation}
Now that we have already showed the base case, it is only left to show that if $b_{kn} = (-1)^{n - k} a_{kn}$ for $k < m$, then $b_{mn} = (-1)^{n - m} a_{mn}$. In general, the $m$-th row results in $m-1$ non-trivial conditions of the form
\begin{equation}
\sum_{k=m-l}^m a_{mk} b_{k m-l} = 0,\quad l = 1,\ldots,m,
\end{equation}
from which it follows
\begin{equation}
b_{m m-l} = -\sum_{k=m-l}^{m-1} a_{mk}b_{k m-l} = -\sum_{k=m-l}^{m-1} (-1)^{k-m+l} a_{mk} a_{k m-l},
\end{equation}
where we have used that, by assumption, $b_{kn} = (-1)^{n - k} a_{kn}$ for $k < m$. Now we plug in the expression for $a_{mn}$ to get
\begin{align}
\label{eq:b_mm-l}
b&_{m m-l} 
=
-\sum_{k=m-l}^{m-1} (-1)^{k-m+l} \frac{m!}{k!(m-k)!} \frac{k!}{(m-l)!(k - m + l)!}\nonumber\\
&\times
\frac{(2m - 1)!!}{(2k - 1)!!} \frac{(2k - 1)!!}{(2m - 2l - 1)!!} q^{m - k} q^{k - m + l} \nonumber\\
&\stackrel{r=k-m+l}{=}
q^l \frac{(2m - 1)!!}{(2m - 2l - 1)!!} \frac{m!}{(m - l)!} \sum_{r = 0}^{l-1} \frac{(-1)^{r+1}}{(l - r)! r!}. 
\end{align}
It remains to show that the last sum is equal to $(-1)^l/l!$. Indeed,
\begin{equation}
\sum_{r=0}^{l-1} \frac{(-1)^{r+1}}{(l-r)!r!} 
=
-\frac{1}{l!} \sum_{r=0}^{l-1} (-1)^r \begin{pmatrix} l \\ r\end{pmatrix}
=
\frac{(-1)^l}{l!},
\end{equation}
where we have used the identity
\begin{equation}
\sum_{r=0}^{k} (-1)^r \begin{pmatrix} n \\ r \end{pmatrix} 
=
(-1)^k \begin{pmatrix} n - 1 \\ k \end{pmatrix}.
\end{equation}
Plugging this into \eqref{eq:b_mm-l} we finally get
\begin{equation}
b_{mm-l}
=
\begin{pmatrix}
m \\
l
\end{pmatrix}
\frac{(2m - 1)!!}{(2m - 2l - 1)!!} (-q)^l
=
\left(U(-q)\right)_{mm-l},
\end{equation}
which closes the proof.

\subsection{Computing $\hat{H}_0$ and $\hat{V}$}
Finally, let us compute $\hat{H}_0$ and $\hat{V}$.  To that end, we first take the derivative of $U(q)$ using \eqref{eq:U_nm}:
\begin{equation}
	\label{eq:dU_nm}
	\partial_y U_{nm}(q)
	=
	\begin{dcases}
		\begin{pmatrix}
			n \\
			m	
		\end{pmatrix}	
		\frac{(2n - 1)!!}{(2m - 1)!!} \left(n - m\right) q^{n-m-1} \partial_y q, \quad &n > m,\\
		0\,, \quad &\text{otherwise}.
	\end{dcases}
\end{equation}
Since $\partial_y U(q)$ is again lower triangular and in addition its diagonal elements are all zero and $U(q)^{-1}$ is lower triangular, too, the product of two, $U(q)^{-1} \partial_y U(q)$, will be lower triangular with zero diagonal elements as well. Hence, $\left[U(q)^{-1} \partial_y U(q)\right]_{nm} = 0$ for $n \leq m$. To get the remaining entries, we simply multiply the two matrices:
\begin{widetext}
\begin{align}
	\left[U(q)^{-1} \partial_y U(q) \right]_{nm} 
	&=
	\partial_y q \sum_{k=m+1}^n \frac{n!}{k!(n-k)!} \frac{k!}{m!(k-m)!} \frac{(2n-1)!!}{(2k-1)!!} 
	\frac{(2k-1)!!}{(2m-1)!!}(k-m)\left(-1\right)^{n-k} q^{n-m-1}\nonumber\\
	&=
	\partial_y q \left(-q\right)^{n-m-1} (m + 1) \frac{(2n - 1)!!}{(2m - 1)!!} \sum_{k=m+1}^{n} (-1)^{k-m-1} 
	\begin{pmatrix}
		n \\
		k
	\end{pmatrix}
	\begin{pmatrix}
		k \\
		m + 1
	\end{pmatrix}
	=
	\partial_y q \left(-q\right)^{n-m-1} (m + 1) \frac{(2n - 1)!!}{(2m - 1)!!} \delta_{n,m+1}\nonumber\\
	&=
	\partial_y q \left(m + 1\right) 	\frac{(2m + 1)!!}{(2m - 1)!!}\delta_{n,m+1}
	=
	\partial_y q n \left(2n - 1\right) \delta_{n,m+1},
\end{align}
where we have used the identity
\begin{equation}
	\sum_{k=m}^n (-1)^{k-m} \begin{pmatrix} k \\m \end{pmatrix} \begin{pmatrix} n \\ k \end{pmatrix} = \delta_{mn}.
\end{equation}
\end{widetext}

\end{appendix}
	
\end{document}